\documentclass[12pt,preprint,showkeys]{revtex4-2}
\usepackage{graphicx}
\usepackage{amsmath}
\usepackage{amsfonts,amssymb}
\usepackage[matrix,arrow,curve,frame,poly,arc]{xy}
\usepackage[english]{babel}

\def\be{\begin{equation}}
\def\ee{\end{equation}}
\def\ba{\begin{eqnarray}}
\def\ea{\end{eqnarray}}

\begin{document}
\bibliographystyle{plainnat}

\title{General vacuum solution of modified $F(G)$-gravity with Gauss-Bonnet term}

\author{Maria V. Shubina}

\email{yurova-m@rambler.ru}

\affiliation{Skobeltsyn Institute of Nuclear Physics\\Lomonosov Moscow State University
\\ Leninskie gory, GSP-1, Moscow 119991, Russian Federation}


\begin{abstract}

In this paper we present a general vacuum solution of the modified Gauss–Bonnet gravity equations for the Friedmann-Lemaître-Robertson-Walker metric. We use an ansatz to reduce the gravitational equations to an ordinary differential ones for function $ F = F(G)$. The solution obtained depends on an arbitrary function and is new. As an example we take an arbitrary function in the form of a power one and analyze the solutions for both $ F(G) $ and the main cosmological physical quantities as scale factor, Hubble rate, the Gauss-Bonnet term and the scalar curvature.

\end{abstract}


\maketitle
\newpage

\section{Introduction}

Without doubting that today General Theory of Relativity (GR) remains the best fundamental theory that describes the gravitational interaction one have to admit that this theory needs modification to describe some observed facts such as the late-time acceleration confirmed in the late $90’s $ \cite{Riess} and the early-time acceleration. The theoretical justification for these and other phenomena is given within the framework of modified theories of gravity that have been widely developed in the last few decades \cite{NO_2008}-\cite{NOO} and references therein. General Relativity is described by the Einstein-Hilbert gravitational action with Lagrange density $ \sqrt{-g} R $, where $ R $ is the Ricci curvature ($ g = \det g_{\mu \nu }$), and among the modified gravity theories the simplest and the most popular modification is the $F(R)$ one. The various forms of $F(R)$ gravity that correctly describe the cosmological dynamic of the Universe both in our time and in early times have appeared in the literature. But in four dimensions along with this modification of GR there are other promising theoretical descriptions such as $ F(G) $ theory of gravity where $ G $ is the Gauss-Bonnet invariant \cite{NO_2005}-\cite{O_2015} and references therein. In these works it was shown in particular that within the framework of the "modified Gauss–Bonnet gravity" \cite{NOO} the late-time acceleration of the universe is described. In the work \cite{O_2015} the connection of $ F(G) $ theory with the bounce cosmology theories is considered.

In this paper we obtain the general vacuum solution of modified Gauss–Bonnet gravity in the Friedmann-Lemaître-Robertson-Walker metric. This solution corresponds to the ansatz that we imposed on the Hubble rate and depends on an arbitrary function. Further we give an example of a solution when this arbitrary function is a power one; we consider different values of the exponent and analyse the expressions obtained for the Hubble rate, the scale factor and e-foldings number.

\section{Models under consideration and field equations}

The modified Gauss–Bonnet gravity $F(G)$ model without matter fields has the action  \cite{NOG_2006}-\cite{NO_2014}:
\be
S = \frac{1}{2k^{2}} \int d^{4}x \sqrt{-\textit{g}} \, (R + F(G)),
\ee
where $ k^{2} = 1/M_{Pl}^{2}$, $ M_{Pl} = 1.22 \times 10^{19} $ GeV, $ \textit{g} $ is the determinant of the metric tensor $ g_{\mu\nu} $, $ G = R^2 - 4R^{\mu \nu} R_{\mu \nu} + R^{\mu \nu \rho \sigma }  R_{\mu \nu \rho \sigma } $ is the Gauss-Bonnet term, $ R = g^{\mu\nu} R_{\mu\nu} $ is the scalar curvature, or the Ricci scalar, $ R_{\mu\nu} $ is the Ricci tensor. Variation of eq.(1) with respect to the metric gives the field equations 
\ba
& R_{\mu \nu}& -  \frac{1}{2} g_{\mu \nu} R - \frac{1}{2}g_{\mu \nu}  F(G) - \big[ -2 R R_{\mu \nu } + 4 R_{\mu \rho } {R_{\nu}}^{\rho} - 
2 {R_{\mu}}^{\rho \sigma \tau} R_{\nu \rho \sigma \tau } + 4 g^{\alpha \rho} g^{\beta \sigma}  R_{\mu \alpha \nu \beta} R_{\rho \sigma}  \nonumber \\
& - & 2 R \, \nabla_{\mu} \nabla_{\nu} + 2 g_{\mu \nu} R \, \square - 4 R_{\mu \nu} \, \square  + 4 {R_{\nu}}^{\rho} \nabla_{\rho} \nabla_{\mu} + 4 {R_{\mu}}^{\rho} \nabla_{\rho} \nabla_{\nu} - 4 g_{\mu \nu} R^{\rho \sigma} \nabla_{\rho} \nabla_{\sigma} \nonumber \\
& + & 4 g^{\alpha \rho} g^{\beta \sigma}  R_{\mu \alpha \nu \beta} \nabla_{\rho} \nabla_{\sigma} \big] F_{G} =  0,
\ea
where $ F_{G}(G)\equiv \dfrac{dF(G)}{dG} $. We will consider a Friedmann–Lemaître–Robertson–Walker (FLRW) metric interval:
\be
ds^{2} = - dt^{2} + (a(t))^{2} \sum \limits_{i = 1, 2, 3} (dx^{i})^{2}.
\ee
Then eqs. (2) take the form:
\ba
& 24 & H^{3} \dot{G}\, F_{G G} - G F_{G} +  F  + 6 H^{2}= 0 \\
& 8 & H^{2} \dot{G}^{2} \, F_{G G G} + \big(16 H \dot{G} ( \dot{H} + H^2) + 8 H^{2} \ddot{G}\big) \, F_{G G} - G F_{G} + F + 4\dot{H} + 6 H^{2} = 0;
\ea
it can be shown that eq. (5) is a consequence of eq. (4) and of expression for the Gauss-Bonnet term $ G = 24 H^2 (\dot{H} + H^2) $. The scalar curvature $ R  =  6 \, (\dot{H} + 2 H^{2})  $ where the Hubble rate $ H = \frac{\dot{a}}{a} $ ($ \dot{ } \equiv \frac{d}{dt}$). Further arguments are similar to those given in the work \cite{Sh_2024} for the $ F(R) $ gravity. Eq. (4) contains differentiation with respect to two variables: $ t $ and $ G $, therefore the writing this equation in terms of single variable $ t $ or $ G $ without concretization of the form of $ H(t) $ (or $ G $) is not possible. 

In this article we choose for the Hubble rate $H$ the similar ansatz as in \cite{Sh_2024} and obtain a exact general solution of eq. (4) for the function $ F(G) $ depending on an arbitrary function $ \Phi(H^2) $. Next we consider in detail the case of function $ \Phi(H^2) \sim (H^2)^{p} $ and obtain different types of solutions for different values of $ p $.

\section{The chosen ansatz and the investigated equation}

We will look for solutions using the cosmological reconstruction technique developed in \cite{NO_2006},\cite{OO_2021}. This technique allows to pass from the cosmological time variable $ t $ to a new e-folding number variable $ N = \ln{\frac{a}{a_{0}}} $. Because  $  \frac{d}{dt} = \sqrt{H^{2}} \frac{d}{dN} $ our ansatz will have the form:
\be
 H^{2} \frac{d H^{2}}{dN} = \Phi(H^{2})
\ee
so that 
\be
G  =  12 \, \big(\Phi(H^{2}) + 2 H^{4}\big).
\ee
Let us introduce the notation $ H^{2} \equiv x $ and rewrite the functions entering the eq. (4) in terms of $ x $. Because
\ba
F_{G}  =  \frac{F_{x}}{G_{x}}, \,\,\, 
F_{GG}  =  \frac{ F_{xx} G_{x} -  F_{x} {G_{xx}}}{(G_{x})^{3}} 
\ea
and $G  =  12 \, (\Phi(x) + 2 x^{2}) $
eq. (4) takes the form:
\be
\frac{2 x \Phi }{\Phi_{x} + 4x} \, F_{xx} - \Big(\frac{2 x \Phi (\Phi_{xx} + 4)}{(\Phi_{x} + 4x)^2} + \frac{\Phi + 2 x^{2}}{\Phi_{x} + 4x}\Big)\, F_{x} + F + 6x = 0.
\ee
In the next section we will obtain a general solution to this equation for an arbitrary function $ \Phi(x) $. By choosing $ \Phi(x) $ we can immediately solve eq. (6) for $ H = H(N) $:
\be
\int\frac{x dx}{\Phi(x)} = N + N_{0},
\ee
where $ N_{0} = const $ and taking into account that $ H = \frac{dN}{dt} $ we obtain $ N(t) $:
\be
\int\frac{dN}{\sqrt{x(N)}} = t + \tilde{t_0}, \,\,\, \tilde{t_0} = const.
\ee

\section{General exact solutions}

Let us first solve the homogeneous equation (9). One can make sure that the function 
\be
F^{(0)}_{1}(x) = c \, \big(2 x^2 + \Phi(x) \big), \,\,\, c = const
\ee
is the first solution to the homogeneous equation (9). Then the general solution is constructed according to the expression \cite{ZP} as
\ba
F^{(0)}(x) = F^{(0)}_{1}(x) \Big( C_1 + C_2 \, \int \frac{\sqrt{x}(\Phi_{x} + 4x)\, e^{\int \frac{xdx}{\Phi}} dx}{(F^{(0)}_{1})^{2}}\Big),
\ea
where the second term is the second solution $ F^{(0)}_{2}(x) $; $ C_1$ and $C_2 $ are constant. To move on to solving the inhomogeneous equation (9) we will use the method of variation of the constant: now $ C_1$ and $C_2 $ are not constant but functions of $ x $. Without loss of generality one can impose the following condition on functions  $ C_1 (x)$ and $C_2 (x)$: $ C_{1 ,x} F^{(0)}_{1} + C_{2 ,x} F^{(0)}_{2} = 0 $ \cite{Elsholtz}. Substituting $F_{i}$ with $ C_i (x) $, $ i = 1, 2 $ into eq. (9) we obtain
\ba
C_1 (x) & = & \int \Big( \frac{1}{c} \int \frac{\sqrt{x} \, (\Phi_{x} + 4x)\, e^{\int \frac{xdx}{\Phi}}\, dx}{(\Phi + 2 x^{2})^2} \Big) \, \frac{3 (\Phi + 2 x^{2})\, e^{- \int \frac{xdx}{\Phi}} }{\sqrt{x} \, \Phi} \, dx + C^{(0)}_1 \nonumber \\ 
C_2 (x) & = & -3c \int \frac{ (\Phi + 2 x^{2})\, e^{- \int \frac{xdx}{\Phi}} }{\sqrt{x} \, \Phi} \, dx + C^{(0)}_2,
\ea
$ C^{(0)}_1 $ and $ C^{(0)}_2 $ are constant. Finally substituting this into eq. (13) for solution of inhomogeneous equation after integrating by parts we obtain the general solution in the form:
\ba
F(x) = c \, \big(2 x^2 + \Phi(x) \big)\,\, \Big[  C^{(0)}_1 + \frac{ C^{(0)}_2 }{c^{2}} \,  \int \frac{\sqrt{x} \, (\Phi_{x} + 4x)\, e^{\int \frac{xdx}{\Phi}}\, dx}{(\Phi + 2 x^{2})^2} \nonumber \\ 
 -  \int  \frac{3}{c} \Big( \int \frac{ (\Phi + 2x^2)\, e^{- \int \frac{xdx}{\Phi}}\, dx}{\sqrt{x} \, \Phi} \Big) \, \frac{\sqrt{x} (\Phi_{x} + 4 x)\, e^{ \int \frac{xdx}{\Phi}} }{(2 x^2 + \Phi)^2}  dx  \Big]. 
\ea
It is obvious that the integrals in this expression will not be taken in elementary functions for most arbitrary $ \Phi(x) $. Next we will give examples of solutions for the simplest form of function $ \Phi(x) \sim x^{p} $ and show that even in this case the analysis of the obtained solutions is difficult for arbitrary $ p $.

\section{$ \Phi(x) = C_{0} x^{p}$}

As follows from eq. (15), for $ \Phi(x) = C_{0} x^{p}$ the general solution has the form:
\ba
F(x) = c \, \big(2 x^2 + C_{0} x^{p} \big)\,\, \Big[  C^{(0)}_1 + \frac{ C^{(0)}_2 }{c^{2}} \,  \int \frac{ \sqrt{x} \, (C_{0} p \, x^{p-1} + 4x)\, e^{\frac{1}{C_0} \int x^{1-p} dx}}{(C_{0} x^{p} + 2 x^{2})^2} dx \nonumber \\ 
 -  \int  \frac{3}{c \, C_{0}} \Big( \int \frac{ (C_{0} x^{p} + 2x^2)\, e^{- \frac{1}{C_0} \int x^{1-p} dx}\, dx}{x^{p + \frac{1}{2}}} \Big) \, \frac{\sqrt{x} (C_{0} p \, x^{p-1} + 4 x)\, e^{ \frac{1}{C_0} \int x^{1-p} dx} }{(2 x^2 + C_{0} x^{p})^2}  dx  \Big]
\ea
and one can immediately see that the solution falls into two "classes": one for $ p = 2 $ and the second for $ p \neq 2 $. Let us start with the case when $ p = 2 $.

\subsection{$ p = 2 $}

For the case $ \Phi(x) = C_{0} x^{2}$ we first obtain explicit expressions for the functions $ x(t) $, the scale factor $ a(t) $ and e-foldings number $ N(t) $:
\ba
x(t) & = & (- \frac{C_{0}}{2}  \,t + t_{0})^{-2}   \\ \nonumber
a(t) & = & a_{0} e^{-N_0} \, \Big| - \frac{C_{0}}{2} \, t + t_{0}\Big|^{-\frac{2}{C_{0}}} \\ \nonumber
N(t) & = & - \frac{2}{C_{0}}\, \ln \Big| - \frac{C_{0}}{2}  \, t + t_{0}\Big|  -  N_0,
\ea
The Gauss-Bonnet term and the scalar curvature have the form:
\ba
G(t) & = & \frac{12(C_{0} + 2)}{(- \frac{C_{0}}{2}  \,t + t_{0})^{4}} \\ 
R(t) & = & \frac{3 (C_{0} + 4)}{(- \frac{C_{0}}{2}  \,t + t_{0})^{2}}
\ea
and one can see what should be $C_{0} \neq -2 $ and $ -4 $.

The similar expressions for $ x $ and $ R $ was obtained in the $ F(R) $-gravity frame. When we consider eq. (17) we can see that for $ C_{0} < 0 $ and for $ t_{0} \geq 0 $ the scale factor is $ a(t) \sim (t + t_{0})^{\frac{2}{|C_{0}|}} $ and it grows monotonically with increasing time. Similarly the Hubble rate, the Gauss-Bonnet term and the scalar curvature are continuous functions for all $ t > 0 $. For $ t_{0} < 0 $ and negative $ C_{0} $ the functions $ H $, $ G $ and $ R $ diverge but this singularity "in the past" may be purely coordinate one. 

For positive $ C_{0}$ the above functions are defined only for $ t < t_{s} $ and have the future singularities at $ t = t_{s} $. Since it is easy to show that $ \kappa ^{2}\rho_{eff} = 3 H^{2} \longrightarrow \infty $ and $ \kappa ^{2} | p_{eff}| = |2 \dot{H} +  3 H^{2}| \longrightarrow \infty  $ ($ \kappa^{2} = 8 \pi G_{N} $, $  G_{N}$ is the Newton’s gravitational constant) at $ t \longrightarrow t_{s} $ this is the Type I (Big Rip) singularity, see \cite{NOO}, \cite{HNOOP} (and references therein) and $ t_{s} $ is the so-called Rip time. This solution describes the universe that ends at the Big Rip singularity in the time $ t_{s} $. 

Let us now move on to solving eq. (9). One can use the general form of solution (16) but it is easier to solve the equation itself which takes the form:
\be
\frac{C_0}{C_0 + 2} \, x^2 F_{xx} - \frac{3 C_0 + 2}{2 (C_0 + 2)} \, x F_{x} + F = - 6x. 
\ee
When the right side of this equation is equal to zero it is the Euler one. The solution to the inhomogeneous eq. (20) has two different forms.

\textbf{\textit{Case 1}:   \textit{$ {C_{0}} = \frac{2}{3} $}}

Eq. (20) becomes
\be
x^2 F_{xx} - 3 \, x F_{x} + 4 \, F = - 24x 
\ee
and its solution is
\be
F(x) = - 24 x + 4 c \, C^{(0)}_1 \, x^2 + \frac{4 \, C^{(0)}_2 }{3c} \, x^2 \, \ln x,
\ee
$C^{(0)}_1 $ and $C^{(0)}_2 $ are constant, and expressed through the Gauss-Bonnet term this function is:
\be
F(G) = - 3 \sqrt{2} \sqrt{G} + \frac{c \, C^{(0)}_1}{8}  \, G + \frac{ C^{(0)}_2 }{48 c} \, G \, \ln \frac{G}{32}.
\ee
It is also interesting to write this function in terms of scalar curvature $ R $. Since $ G = \frac{8}{49} R^2 $ we obtain:
\be
F(R) = - \frac{12}{7} R + \frac{4 c \, C^{(0)}_1}{196}  \, R^2 + \frac{ C^{(0)}_2 }{147 c} \, R^2 \, \ln \frac{R}{14};
\ee
the comparison with the solution for a similar form of the function $ \Phi $ but within the framework of $ F(R) $ gravity shows that for $ {C_{0}} = \frac{2}{3} $ both the equations and the resulting solutions have different forms \cite{Sh_2024}.

\textbf{\textit{Case 2}:   \textit{$ {C_{0}} \neq \frac{2}{3} $}}

Solving the inhomogeneous eq. (20) using the method of variation of constants we obtain for the functions $ C_1(x) $ and $ C_2(x) $:
\ba
C_1 (x) & = & - \frac{12 C_0}{c (2 - 3 C_0)} \, \frac{1}{x} + C^{(0)}_1 \nonumber \\
C_2 (x) & = & - \frac{6 c  (C_0 + 2)}{C_0 (C_0 - 2)} \, x^{\frac{C_0 - 2}{2 C_0}} + C^{(0)}_2
\ea
and the general solution has the form:
\be
F(x) = \frac{12  (C_0 + 2) }{C_0 - 2} \, x + \frac{c  (C_0 + 2) \, C^{(0)}_1 }{C_0}  \, x^2 + \frac{4 C_0^2 \, C^{(0)}_2 }{c (2 - 3 C_0)} \, x^{\frac{C_0 + 2}{2 C_0}}.
\ee
Let us express this function in terms of $ G $:
\be
F(G) = \frac{2 \sqrt{3 (C_0 + 2) \, G} }{C_0 - 2}  + \frac{c \, C^{(0)}_1 }{12 C_0}  \, G + \frac{2 C_0^2 \, C^{(0)}_2 }{ \sqrt{3} c (2 - 3 C_0)} \, \Big( \frac{G}{C_0 + 2} \Big)^{\frac{C_0 + 2}{4 C_0}}.
\ee
Because $ x = \frac{R}{3 (C_0 + 4)} $ it is easy to write $ F(R) $:
\be
F(R) = \frac{4  (C_0 + 2) }{(C_0 - 2)(C_0 + 4)} \, R + \frac{c  (C_0 + 2) \, C^{(0)}_1 }{9 C_0 (C_0 + 4)^2}  \, R^2 + \frac{4 C_0^2 \, C^{(0)}_2 }{c (2 - 3 C_0)} \, \Big( \frac{R}{3(C_0 + 4)} \Big)^{\frac{C_0 + 2}{2 C_0}}
\ee
and just as in the previous case this type of function cannot be obtained within the $ F(R) $ gravity framework with a power-law choice of $ \Phi(x) $ \cite{Sh_2024}. 

In the work \cite{O_2015} the solutions for $ H \sim (t - t_s)^\alpha $ and $ F(G) \sim \Sigma C(k) G^{k(\alpha)} $, $ k(\alpha) = 1, \frac{\alpha}{3\alpha - 1}$ and $ \frac{2\alpha}{3\alpha - 1} $ were obtained and analyzed in the context of bounce cosmology. In our case $ \alpha = - 1 $ and $ k = 1, \frac{1}{2}$ and $ \frac{C_0 + 2}{4 C_0} $ where the first two values are the same as in \cite{O_2015} but the value $ \frac{1}{4} $ cannot be obtained. The time dependence of Gauss-Bonnet term $ G(t) \sim (t - t_s)^{-4} $ (see eq. 18) coincides with the one obtained in the work \cite{O_2015}.

\subsection{$ p \neq 2 $}

As with $ p = 2 $ we again start with expressions for the functions $ x $. From eq. (10) it follows that 
\be
x(N) = \Big( C_0 (2 - p) \, (N + N_0) \Big)^{\frac{1}{2-p}}
\ee
however eq. (11) takes the form
\be
\int \Big( C_0 (2 - p) \, (N + N_0) \Big)^{ - \frac{1}{2(2-p)}} dN = t + \tilde{t_0}
\ee
and one can see that again two cases arise already for values $ p \neq 2 $.

\textbf{\textit{Case 1}:   \textit{$ p \neq \frac{3}{2} $}}

In this case the expressions for the functions $ x (t)$, the scale factor $ a(t) $ and e-foldings number $ N(t) $ are as follows:
\ba
x(t) & = & \Big( \frac{C_{0} (3 - 2p)}{2}  \,t + t_{0} \Big)^{\frac{2}{3-2p}}   \\ \nonumber
a(t) & = & a_{0} e^{\frac{1}{C_{0} (2 - p)} \, \Big( \frac{C_{0} (3 - 2p)}{2}  \,t + t_{0} \Big)^{\frac{2 (2-p)}{3-2p}} - N_0}  \\ \nonumber
N(t) & = & \frac{1}{C_{0} (2 - p)} \, \Big( \frac{C_{0} (3 - 2p)}{2}  \,t + t_{0} \Big)^{\frac{2 (2-p)}{3-2p}} - N_0,
\ea
where $ t_0 = const $. 

Let us now consider what types of finite time singularities at $ t = t_s  $ the obtained solutions have and under what conditions. Since $  \frac{C_{0} (3 - 2p)}{2}  \,t + t_{0} > 0 $ and for the finite time singularity to exist there must be $ \frac{C_{0} (3 - 2p)}{2} \,t + t_{0} \sim t_{s} - t $ it follows that $ C_{0} (3 - 2p) < 0 $. Further following the classification of singularities given in the works \cite{NOO}, \cite{HNOOP} (and references therein) we will again consider the behaviour of $ \kappa ^{2}\rho_{eff} $ and $ \kappa ^{2} | p_{eff}| $ at $ t \longrightarrow   t_s $. It can be shown that depending on the values of $ p $ all 4 types of singularities are realized. So for $ p < 1 $ ($ C_0 < 0 $) we have the Type II singularity because $ a \longrightarrow a_s $, $ \rho_{eff} \longrightarrow \rho_{s} $ where $ \rho_{s} = 0 $ and $ | p_{eff}| \longrightarrow \infty $. For $ 1 < p < \frac{3}{2} $ ($ C_0 < 0 $) the Type IV singularity is realized: $ a \longrightarrow a_s $, $ \rho_{eff} \longrightarrow 0 $ and $ | p_{eff}| \longrightarrow 0 $. When $ \frac{3}{2} < p < 2 $ ($ C_0 > 0 $) $ a \longrightarrow \infty $, $ \rho_{eff} \longrightarrow \infty $ and $ | p_{eff}| \longrightarrow \infty $ and we have the Type I singularity. Finally for $  p > 2 $ ($ C_0 > 0 $) $ a \longrightarrow a_s $, $ \rho_{eff} \longrightarrow \infty $ and $ | p_{eff}| \longrightarrow \infty $ and this is the Type III singularity. 

The Gauss-Bonnet term and the scalar curvature have the form:
\ba
G(t) & = & 12 \Big(C_{0} \big(\frac{C_{0} (3 - 2p)}{2}  \,t + t_{0} )^{\frac{2p}{3-2p}} + 2 ( \frac{C_{0} (3 - 2p)}{2}  \,t + t_{0} \big)^{\frac{4}{3-2p}}    \Big) \\ 
R(t) & = & 3 \Big(C_{0} \big(\frac{C_{0} (3 - 2p)}{2}  \,t + t_{0} )^{\frac{2(p-1)}{3-2p}} + 4 ( \frac{C_{0} (3 - 2p)}{2}  \,t + t_{0} \big)^{\frac{2}{3-2p}}    \Big).
\ea
Let us rewrite eq. (16) for the considered values of $ p $ as:
\ba
F(x) & = & c \, \big(2 x^2 + C_{0} x^{p} \big)\,\, \Big[  C^{(0)}_1 + \frac{ C^{(0)}_2 }{c^{2}} \, 
\Big( - \frac{\sqrt{x} \, e^{\frac{x^{2-p}}{C_0 (2-p)}}}{C_0 x^p + 2x^2} + \frac{1}{2C_0} \int x^{ - p - \frac{1}{2} } \, e^{\frac{x^{2-p}}{C_0 (2-p)}} dx \Big)  \\ 
& - & \frac{3}{c C_0} \int \Big( \int  (C_{0} x^{- \frac{1}{2}} + 2x^{-p + \frac{3}{2}})\, e^{- \frac{x^{2-p}}{C_{0} (2-p)}}\, dx \Big) \, \Big( - \frac{\sqrt{x} \, e^{\frac{x^{2-p}}{C_0 (2-p)}}}{C_0 x^p + 2x^2} + \frac{1}{2C_0} \int x^{ - p - \frac{1}{2} } \, e^{\frac{x^{2-p}}{C_0 (2-p)}} dx \Big)   dx
\Big]. \nonumber
\ea
Consider now the obtained solution in more detail. As one can see we have three types of integrals in parentheses, see eq. (34): $ \int x^{q_{1, 2}} e^{ - \frac{x^{2-p}}{C_0 (2-p)}} dx $ and $\int x^{q_{3}} e^{ \frac{x^{2-p}}{C_0 (2-p)}} dx $ where $ q_1 =  \frac{3}{2} - p $, $ q_2 = - \frac{1}{2} $, $ q_3 = - p - \frac{1}{2} $. Let us move under the integrals to a new variable $ t =  - \frac{x^{2-p}}{C_0 (2-p)} $ and let $ t > 0 $. Then the integrals take the form: $ \int t^{\frac{q_{i} + p - 1}{2 - p}} e^{\pm t} dt  $, $ i = 1, 2, 3 $; briefly they look like $ \int t^{\tilde{q}} e^{\pm t} dt $, $ \tilde{q}\lessgtr 0 $ and they can be expressed through an incomplete gamma function \cite{BE_1953}. For definiteness let us consider the case $ C_{0} (3 - 2p) < 0 $. For $ p < \frac{3}{2} $ one can see from eq. (31) that for $ 0 < t < t_{s} $ $ x $ decreases from $ x_s = {t_s}^{\frac{2}{3-2p}} $ to $ 0 $. For $ p > \frac{3}{2} $ $ x $ increases from $ x_s $ to $ \infty $. Then the indefinite integral $ \int t^{\tilde{q}} e^{\pm t} dt  = \int t^{\tilde{q}} \Sigma_{n=0}^\infty \frac{(\pm t)^n}{n!} dt = \Sigma_{n=0}^\infty \int t^{\tilde{q}} \frac{(\pm t)^n}{n!} dt = \Sigma_{n=0}^\infty \frac{ (\pm 1)^{n} \tilde{x}^{\tilde{q} + n + 1}}{n! (\tilde{q} + n + 1)} $ where $ \tilde{x} = - \frac{{x}^{2-p}}{C_0 (2-p)} > 0 $ and $ \int \Sigma = \Sigma \int $ because the exponential series converges uniformly. Some of these integrals can be expressed in terms of the lower ($ x < x_s $) and upper ($ x > x_s $) incomplete gamma function, but the lower one is defined as $ \int_0^{\tilde{x}}... $ ($ \tilde{q} + 1 > 0 $) and one can see that for $ q_{3} $ the series diverges at $ x=0 $ for $ \frac{1}{2} < p < \frac{3}{2} $. Similarly it can be shown that for the upper incomplete gamma function  $ \int_{\tilde{x}}^{\infty}... $ the series diverges for $ q_{3} $ at $ \frac{3}{2} < p < 2 $. 

Let us consider the case when the integrals can be taken in elementary functions. We will impose a condition on $ p $, namely, let $ \frac{q_{i} + p - 1}{2 - p} $ be an integer. So, $ \frac{q_{i} + p - 1}{2 - p} = n $, $ n \in Z $. Then $ p = 2 - \frac{1}{2n} $ and $ n \neq 0 $. Further $ 2 p - 3 = \frac{m}{n} $, $ m \in Z $ and we have $ m = n-1 $. And one more condition for $ n $ and $ m $ is: since we are now considering $  p \neq \frac{3}{2} $ we have $ n \neq 1 $ and $ m \neq 0 $. In this case the integrals are taken exactly and the antiderivatives are presented in the form of finite sums of powers of $ t $, and when the degree of $ t $ is negative the integral exponent is also contained. Thus we can consider two cases: $ n > 1 $ ($ m > 0 $) and $ n < 0 $ ($ m < -1 $). In the first case taking into account the condition $ C_0 (2-p) < 0 $ we obtain that $ \frac{3}{2} < p < 2 $, $ C_0 < 0 $ and we can conclude that the finite time singularity is not  realized. In the second case we have $ p > 2 $ and $ C_0 > 0 $, hence the Type III singularity may occur. If we choose $ t = \frac{x^{2-p}}{C_0 (2-p)} > 0 $ then we can obtain the finite time singularity of Type I. However with such a choice of $ p(n) $ for integer $ n $ we will not be able to obtain the singularities of Types II and IV.

Omitting the technical details of the calculation we present the solution for $ n > 1 $: 
\ba
F(x) & = & c \, \big(2 x^2 + C_{0} x^{2 - \frac{1}{2n}} \big)\,\, \Big[  C^{(0)}_1 + \frac{ C^{(0)}_2 }{c^{2}} \, 
\Big( - \frac{\sqrt{x} \, e^{\frac{2n }{C_0}\, x^{\frac{1}{2n}}}}{C_0 x^{2 - \frac{1}{2n}} + 2x^2} \\
& + & \frac{C_0}{2c^2} \, e^{\frac{2n }{C_0}\, x^{\frac{1}{2n}}} \, \Sigma_{\kappa = 2}^{3n}\dfrac{(-1)^{\kappa} \,  \big( - \frac{2n}{C_0}\big)^{ \kappa -1}  \big( x^{\frac{1}{2n}} \big)^{-3n + \kappa -1} }{(3n-1)(3n-2)...(3n-\kappa+1)} \nonumber \\
& - & \frac{C_0}{2c^2} \big( - \frac{2n}{C_0} \big)^{3n}\dfrac{(-1)^{3n+1} Ei  \big( \frac{2n}{C_0} x^{\frac{1}{2n}} \big)}{(3n-1)(3n-2)...1}\Big) \nonumber \\
& - & \dfrac{6}{c (C_0 x^{2 - \frac{1}{2n}} + 2x^2)} \, \Big( x + 2 \, \Sigma_{\kappa = 1}^{n} (-1)^{\kappa} n (n-1)...(n-\kappa+1) \big( - \frac{C_0}{2n} \big)^{\kappa} \, x^{1 - \frac{\kappa}{2n}} \Big) \nonumber \\    
& + & \frac{6 C_0 2^{2n} n }{c} \Big( \ln \big(|2x^{\frac{1}{2n}} + C_0 |^{a} \, (2x^{\frac{1}{2n}})^{b_{1}} \big)  -  \Sigma_{\kappa = 2}^{2n} \dfrac{b_{\kappa} \, (2x^{\frac{1}{2n}})^{1-\kappa}}{\kappa - 1} \Big) \nonumber \\
& + & \frac{ 6 C_0 2^{2n+1} n}{c } \, \Sigma_{\kappa = 1}^{n} (-1)^{\kappa} n(n-1)...(n-\kappa +1) \big( -\frac{C_0}{n} \big)^{\kappa} \big( 1 - \frac{\kappa}{2n} \big)  \nonumber \\
& \times & \Big( \ln \big(|2x^{\frac{1}{2n}} + C_0 |^{\tilde{a}} \, (2x^{\frac{1}{2n}})^{\tilde{b_{1}}} \big)  -  \Sigma_{s = 2}^{2n+\kappa} \dfrac{\tilde{b_{s}} \, (2x^{\frac{1}{2n}})^{1-s}}{s - 1} \Big) \Big]. \nonumber
\ea
Here $ a $, $b $, $\tilde{a} $ and $ \tilde{b} $ are the coefficients of the expansion of a proper rational fraction into the sum of simple fractions: $ \frac{1}{z(z-C_0)^Q} = \frac{a}{z} + \frac{b_{1}}{(z-C_0)} + ... +  \frac{b_{Q}}{(z-C_0)^{Q}}$, where $ Q = 2n $ and $ Q = 2n+\kappa $.

Rewriting eqs. (32)-(33) as
\ba
G(x) & = & 12(2 x^2 + C_0 x^{2 - \frac{1}{2n}}) \nonumber \\
R(x) & = & 3(4 x + C_0 x^{1 - \frac{1}{2n}})
\ea
one can see that it is impossible to express $ x(G) $ or $ x(R) $.

\textbf{\textit{Case 2}: \textit{$ p = \frac{3}{2} $}}

For $ p = \frac{3}{2} $ the expressions for the functions $ x (t)$, the scale factor $ a(t) $ and e-foldings number $ N(t) $ take the form:
\ba
x(t) & = & C_1 e^{C_0 t}  \\ \nonumber
a(t) & = & a_{0} e^{\frac{2 \sqrt{C_1}}{C_{0}} \, e^{\frac{C_0}{2} t} - N_0}  \\ \nonumber
N(t) & = & \frac{2 \sqrt{C_1}}{C_{0}} \, e^{\frac{C_0}{2} t} - N_0,
\ea
where $ C_1 $ is a positive constant. The expression for function $ F(x) $ will be obtained from eq. (35) if we substitute $ n = 1 $. By writing eq. (36) for the Gauss-Bonnet term and the scalar curvature as
\ba
G(H) & = & 12(2 H^4 + C_0 H^{3}) \nonumber \\
R(H) & = & 3(4 H^2 + C_0 H).
\ea
It can be seen that it is possible to obtain an explicit expression for $ x(G) $ and then express $ F(G) $ but this will be very cumbersome.

\section{Conclusion}

In this paper we consider the modified gravity equations with the Gauss-Bonnet term in the Friedmann-Lemaître-Robertson-Walker metric. We impose the ansatz on the Hubble rate similar to one introduced in \cite{Sh_2024} for the $ F(R) $ modified gravity. This ansatz depends on an arbitrary function and allows to write the gravitational field equations in terms of a single variable. Thus we obtain a general vacuum solution of  the modified Gauss–Bonnet gravity equations corresponding to the chosen ansatz. Next we give an example of the obtained solution by setting an arbitrary function to a power one. We show that even with such a simple choice the general solution consists of several branches corresponding to different values of the exponent; we consider these exponents and analyse the expressions obtained for the Hubble rate, the scale factor, the scalar curvature and the Gauss-Bonnet term.

\end{document}